\documentclass[conference]{IEEEtran}
\IEEEoverridecommandlockouts
\usepackage{cite}
\usepackage{amsmath,amssymb,amsfonts}
\usepackage{algorithmic}
\usepackage{graphicx}
\usepackage{textcomp}
\usepackage{amsmath,bm}
\usepackage{xcolor}
\usepackage{geometry}
\usepackage{float}
\usepackage{stfloats}
\usepackage{xcolor}
\usepackage{enumitem}
\newcommand{\subscript}[2]{$#1 _ #2$}
\usepackage{amsthm}
\theoremstyle{definition}
\newtheorem{definition}{Definition}[section]

{
      \theoremstyle{plain}
      
  }
\usepackage{colortbl,booktabs}
\usepackage{subfig}
\usepackage{caption}
\def\BibTeX{{\rm B\kern-.05em{\sc i\kern-.025em b}\kern-.08em
    T\kern-.1667em\lower.7ex\hbox{E}\kern-.125emX}}
\begin{document}

\title{Economic Model Predictive Control of Water Distribution Systems with Accelerated Optimization Algorithm
}




\author{Saskia Putri$^{1}$, Faegheh Moazeni$^{2}$, Javad Khazaei$^{3}$
\\
\textcolor{blue}{This work has been submitted to the EWRI Congress 2024 for possible publication.} \\
\textcolor{blue}{Copyright may be transferred without notice, after which this version may no longer be accessible.}
\thanks{The authors are with the Rossin college of engineering and applied science at Lehigh University, Bethlehem, PA 18015, USA. (Emails: 
        {\tt\small sap322@lehigh.edu, moazeni@lehigh.edu,  khazaei@lehigh.edu)}}%
}
\maketitle

\begin{abstract}
Model predictive control (MPC) has emerged as an effective strategy for water distribution systems (WDSs) management. However, it is hampered by the computational burden for large-scale WDSs due to the combinatorial growth of possible control actions that must be evaluated at each time step. Therefore, a fast computation algorithm to implement MPC in WDSs can be obtained using a move-blocking approach that simplifies control decisions while ensuring solution feasibility. This paper introduces a least-restrictive move-blocking that interpolates the blocked control rate of change, aiming at balancing computational efficiency with operational effectiveness. The proposed control strategy is demonstrated on aggregated WDSs, encompassing multiple hydraulic elements. This implementation is incorporated into a multi-objective optimization framework that concurrently optimizes water level security of the storage tanks, smoothness of the control actions, and cost-effective objectives. A fair comparison between the proposed approach with the non-blocking Economic MPC is provided.
\end{abstract}

\vspace{+0.1cm}
\section{INTRODUCTION}
\subsection{Motivation}
A water distribution system (WDS), mainly applied in a large-scale water system, is a sophisticated network that incorporates pumps for water conveyance, valves for flow and water pressure control, and a reservoir to manage demand fluctuation and emergencies \cite{DWDS}. This system plays a vital role in developing smart and resilient urban cities \cite{moazeni2021sequential}. In addition to being a complex critical infrastructure system, WDS is considered a highly nonlinear and constrained system due to the pumps' operation and friction of the pipes \cite{berkel2018modeling,pour_economic_2019}. Therefore, providing optimal control of WDSs is crucial for delivering high-quality and reliable service.  

\subsection{State of the art}
Given the significance of achieving a secure and stable water supply, a robust and multivariate-based control strategy is essential \cite{fambrini_modelling_2009}. Model predictive control (MPC) has emerged as a promising approach \cite{mayne_model_2014,chen1998quasi,castelletti2023model} offering an optimal control approach tailored to WDSs characteristics. These include multi-variate, nonlinear, complex, and interconnected dynamics of WDS assets \cite{tung2020survey}. A set of constraints is obtained by including the physical limitation of the WDSs' elements to restrict the input and state variables, while the cost function is derived based on an optimal control formulation over a finite horizon. 

MPC in WDSs is often utilized to obtain optimal solutions from a multi-objective optimization, including water level security of the tanks, operational costs, smoothness of the control actions to expand instruments lifespan while satisfying water demand \cite{wang_receding_2020,leirens2010coordination,pour_economic_2019}. MPC in WDSs have been developed using linear prediction model \cite{trapiello2023reconfiguration,pour_economic_2019,baunsgaard2016mpc} and nonlinear prediction models \cite{verheijen_efficient_2022,guo2022optimal,wang2020receding}.

Nonetheless, MPC often entails a computational burden when the system is complex, nonlinear, and has multiple inputs and multiple outputs \cite{giovanelli2020move,son2020move}, which is a concern for WDSs. One common method for accelerating MPC optimization is via the move-blocking (MB) strategy, which allows the optimization problem to be tractable. The objective is to confine the degrees of freedom (DoF) of the control inputs in a manner that prevents continuous variation of the control signal \cite{schwickart2016flexible,cagienard2007move}. Commonly, the DoF is reduced by setting the control horizon (number of control moves to be optimized at control interval $k$, $N_c$) to be strictly less than the prediction horizon ($N_p$). Hence, at time step $N_c+1 \to N_p$, the inputs are held constant (blocked inputs). 

However, this approach is only valid when the set point signal is static \cite{schwickart2016flexible}, which is not the case for WDS with demand variation. Furthermore, fixing the input after $N_c$ moves may lead to control performance degradation \cite{cagienard2007move}. Particularly, the distribution of the input blocks over the prediction horizon is necessary to enhance controller performance \cite{son2021move}. Therefore, a blocking interval and position of the move blocking approach are essential to consider to achieve greater flexibility. However, the solution also involves disadvantages where feasibility and compliance with constraints may not be obtained \cite{cagienard2007move}. 

Various studies have delved into investigating the move-blocking techniques \cite{chen2020efficient,li2014fast}. For instance, a move-blocking strategy coupled with constraint-set compression was applied in an automotive system by \cite{li2014fast}. In \cite{chen2020efficient} the blocking technique was utilized to a multiple shooting-based nonlinear MPC to control an inverted pendulum and driving simulator. In another study done by \cite{rossi2023long}, a combination of move blocking scheme and restriction to the candidate solutions were utilized to a grid-tied converter. Nonetheless, most of the studies were applied to a benchmark model and have not been applied to WDSs. 

\subsection{Contribution of this study}
To address the existing challenges associated with reducing the computational load of MPC for WDS operation, the main contribution of this paper is developing a computationally efficient MPC algorithm. The algorithm leverages the piecewise linear move-blocking technique. The core concept centered on yielding a least restrictive move blocking via linear interpolation to the blocked control inputs and predefined the blocking length and distribution, to ensure the optimality. Nonlinear objective functions with time-varying penalty weights in the MPC control-oriented problem formulation for flow-based WDSs are also formulated. It is mainly aimed at reducing the pump's energy consumption concerning time-varying electricity costs. 

\subsection{Proposed model layout}
\begin{figure*}[t!]
    \centering
    \includegraphics[width=1\textwidth]{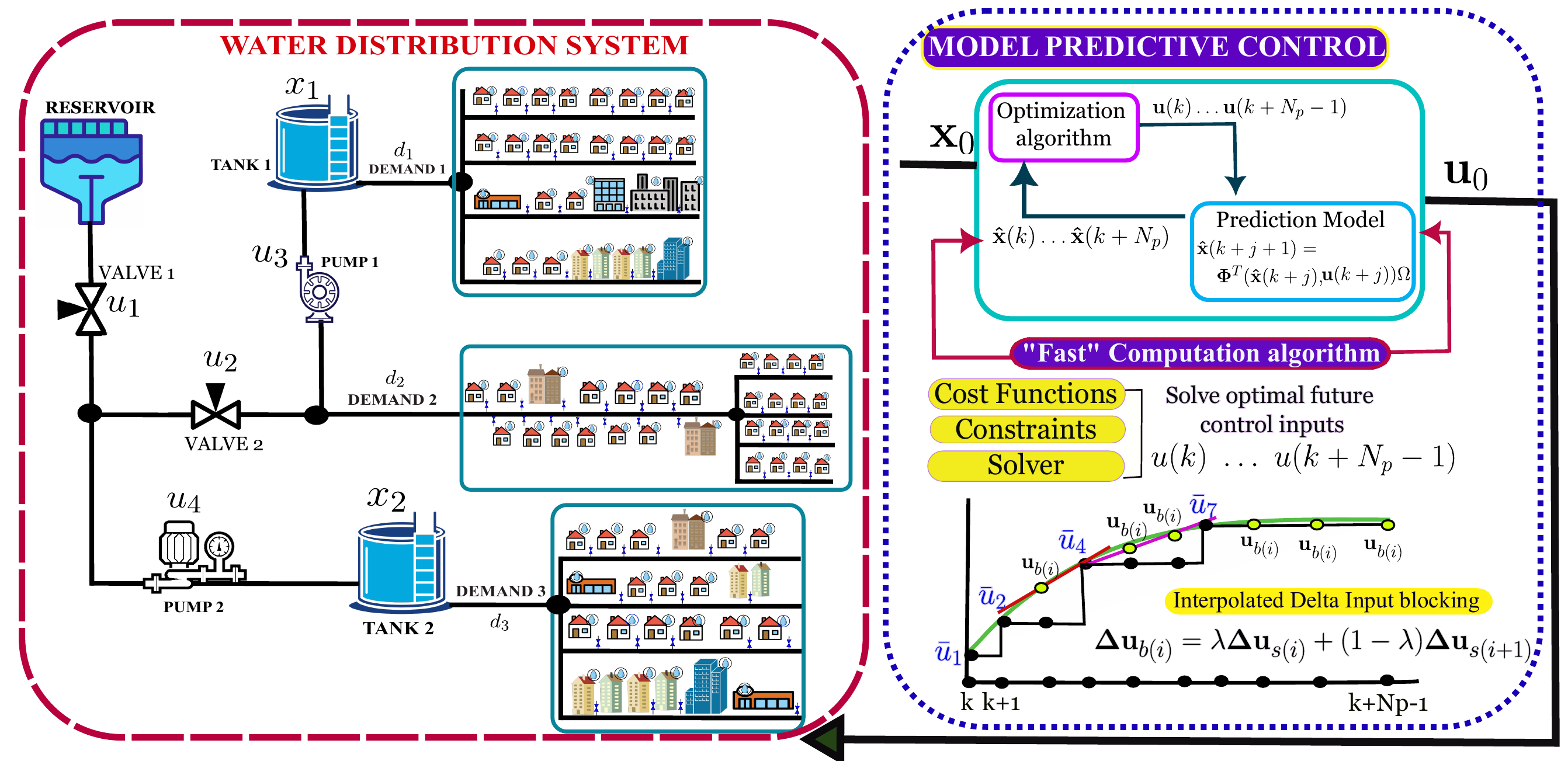}
    \caption{Structure of the proposed fast control algorithm for demand-driven WDS operation}
    \label{layall}%
\end{figure*}

An overview of the proposed least-restrictive control algorithm for demand-driven WDS operation is illustrated in Fig.~\ref{layall}. A conceptual aggregated WDS used in this study features one reservoir, two storage tanks, two pumps, two valves, and some one-directional pipes. The system is designed for an aggregated water demand. It is assumed that all main hydraulic components are installed with smart metering devices. Here, the tanks' water levels ($x_i$) are collected and set as the initial condition for the control-oriented optimization under the fast MPC strategy. The main difference in this study compared to the state-of-the-art MPC is that instead of solving $n$-th control sequence subject to long prediction intervals, here we will use the modified blocking strategy to limit the control input sequence while still guaranteeing convergence and minimum error. The optimal control input generated from MPC is then delivered to the WDS operation, ensuring cost-effective operation and demand compliance. 

\subsection{Structure of the paper}
The rest of the paper is structured as follows. The dynamics of the aggregated WDS including the WDS objectives required for the optimization formulation are presented in Section~\ref{sec:WDS}. Section~\ref{sec:mpc} describes the problem formulation of MPC with the proposed move-blocking strategy while Section~\ref{sec:casestudies} presents time-series validation of the proposed method. Section~\ref{sec:conclusion} concludes the paper. 

\section{MODELING OF WATER DISTRIBUTION SYSTEMS}\label{sec:WDS}
In this study, the mathematical modeling of the aggregated WDS is derived from the governing hydraulic laws. 
\subsubsection{Storage tanks}
WDS' dynamic is typically associated with the storage tanks of the network, which can be expressed as follows \cite{walski_advanced_2003}:
\begin{align}
    \frac{dh_n}{dt} = \frac{1}{A_n}\left(\sum q_{in}(t) - \sum q_{nj}(t)\right) \label{tank1}
\end{align}
where $h_n$ represents the water levels of the $n$-th tank in $m$, $A_n$ as the cross-sectional area of the storage tanks in $m^2$, $q_{in}$ and $q_{nj}$ denote the $i$-th inflow and $j$-th outflows of the $n$-tank, respectively in $m^3/hr$. 

The tank's physical capacity is expressed as:
\begin{align}
    &\underline{h} \leq h_{n}(t)\leq \overline{h} \label{tankLIM}
\end{align}
where $\underline{h}$ and $\overline{h}$ express the minimum and maximum threshold of the water level in $m$, respectively.

\subsubsection{Actuators}
Actuators in this study include pumps and valves, which are regarded as the control inputs.  Constraint associated with the actuators' physical limitation is expressed as:\par
\begin{equation}
    \underline{q}_{i} \leq q_{i,m}(t)\leq \overline{q}_{i} \label{tank3}
\end{equation}
where $\underline{q}_{i}$ and $\overline{q}_{i}$ express the minimum and maximum of the flow rates of the $m$th pumps and valves in $\frac{m^3}{s}$, respectively. 

In addition, nonlinearity in the WDS is often associated with pumps in water distribution systems. There exists a nonlinear relationship between the pressure head, efficiency, and flow rates of pumps which can be computed as follows \cite{walski_advanced_2003}:
\begin{align}
    H_p(t) &= a_{p}(q_{pi}(t))^2+ b_{p}q_{pi}(t)+c_{p} \label{tank4} \\
    \eta_p(t) &= a_{i}(q_{pe}(t))^2+ b_{i}q_{pi}(t)+c_{i} \label{tank5}
\end{align}
where at time $t$, $H_p(t)$ gives the pressure head of the $i$-th pump in $m$, $\eta_p(t)$ denotes the efficiency of the $i$-th pump in $\%$, while $a(\cdot),b(\cdot),c(\cdot)$ are the pump's specific coefficients to compute the pressure head of the $i$-th pump and the efficiency of the $i$-th pump, which can be obtained from the pumps' manufacture.
\subsubsection{Nodes}
Assuming no leakage, flow balances of the nodes are designed as the equality constraint such that \cite{walski_advanced_2003}:
\begin{align}
    d_j (t) = \sum q_{ij} (t) - \sum q_{ji} (t)\label{tank2}
\end{align}
where at time $t \in \mathbb{Z_{+}}$, $d_j$ is the demand of node $j$, $q_{ij}$ are the flows from node $i$ to node$j$ while $q_{ji}$ are the flows from node $j$ to node $i$, all variables' units are in $m^3/hr$.

\section{FAST COMPUTATION OF MODEL PREDICTIVE CONTROL}\label{sec:mpc}
\subsection{Prediction model}
The overall dynamic model of the WDS in Section~\ref{sec:WDS} can be represented in a state-space form as follows:
\vspace{-0.1cm}
\begin{equation}
\vspace{-0.1cm}
    \dot{\bm{x}} = \mathbf{f}(\bm{x}(t),\bm{u}(t),\bm{d}(t)), \quad \bm{x}(0) = \bm{x}_0 \label{wdss}
\end{equation}
where $\mathbf{x} =  [h_1 \ h_2]^T \in \mathbb{R}^{n_x}$ is the state vector, $\mathbf{u} = [q_{v1} \ q_{v2} \ q_{p1} \ q_{p2}]^T \in \mathbb{R}^{n_u}$ is the input vector, and $\mathbf{d} = [q_{d1} \ q_{d2} \ q_{d3}]^T \in \mathbb{R}^{n_d}$ is the measured disturbance vector. 

Integrating the continuous state space model in Eq.~\eqref{wdss} with fourth-order Runge-Kutta \cite{chapra2010numerical} over a sampling period $\delta t$, discrete time representation of the WDS's dynamics is expressed as follows: 
\vspace{-0.1cm}
\begin{align}
\vspace{-0.1cm}
    \bm{x}(k+1) &= \mathbf{f}_d(\bm{x}(k),\bm{u}(k),\bm{d}(k)) \nonumber \\
    &= \bm{x}(k) + \int_{k}^{k + \delta t} \mathbf{f}(\bm{x}(\tau),\bm{u}(\tau),\bm{d}(\tau)) d\tau
 \label{disdyn}
\end{align}
\vspace{-0.5cm}

The prediction model can be developed using the discretized state space model in~Eq.~\eqref{disdyn} over the prediction horizon ($N_p  \in \mathbb{N}^+ $), expressed as follows:
\begin{align}
    \bm{x}(k+j+1) &= \mathbf{f_d}(\bm{x}(k+j),\bm{u}(k+j),\bm{d}(k+j)) \nonumber \\ & \quad \quad \forall j \in \{0,\dots,N_p-1\} \label{nmpc_pred}
\end{align}
where $k = k_0 \to T -1$ represents the duration of the simulation time from the initial ($k_0$) to the final simulation time (T), $j$ is the sampling time-step over $N_p$.

\subsection{Multi-objective WDS control formulation:} 
A multi-objective-based model predictive control is utilized in this paper to manage the WDS optimally, as follows \cite{pour_economic_2019,wang_non-linear_2017}:
\begin{align}
    J = \sum_{j=0}^{N_p-1} \left(w_1l_e (k+j) + w_2l_s (k+j) + w_3l_{\Delta u_i} (k+j)\right) \label{tank7}
\end{align}
where:
\begin{enumerate}[label=(\subscript{O}{{\arabic*}})]
    \item Economic and reliable water supply: power consumption of the pump, together with the time-varying electricity cost, is incorporated into the objective function to ensure a cost-effective operation of pumps. 
    \begin{align}
        l_e(k+j):= \left|\frac{Hp(k)B\psi(k)}{\eta(k)}u_{i}(k)\right| \forall k \in \{1,\hdots, T\} \label{tank8}
    \end{align}
    where $B=\dfrac{\gamma}{1000}$ is a constant for the pump with $\gamma$ representing the specific weight of water ($9.810 \ N/m^{3}$) and at time instant $k$, $Hp(k)$, $\eta(k)$ and $\psi(k)$ are the time-varying pump's pressure head, pump's efficiency, and electricity tariff of the pump. 
    \item Safety water level: to maintain appropriate water levels in tanks for emergency purposes, the controller is regulated to minimize the deviation between the actual tanks' water levels and the desired level. 
    \begin{align}
        l_s(k):= \|x_i(k) - x_i^{ref}\|^2_2 \label{tank9}
    \end{align}
   where $x_i^{ref} \in \mathbb{R}$ denotes the desired level of the the $i$-th tanks. However, this objective is set to be of the lowest priority as the goal of the WDS control is to meet the demand at all times. 
    \item Actuator's smoothness: In this study, pumps and valves are the actuators of the WDS and thereby providing smooth control actions is necessary to increase the equipment's lifespan by protecting it from sudden changes in demand. 
    \begin{align}
        l_{\delta u_i}:= \|\Delta u_i\|^2_2 \label{tank10}
    \end{align}
    \item $w(.)$ denotes the weighted factor to determine the priority of each objective. 
\end{enumerate}

\subsection{Input Profiles}
\begin{figure}[t!]
    \centering
    \includegraphics[width=1\columnwidth]{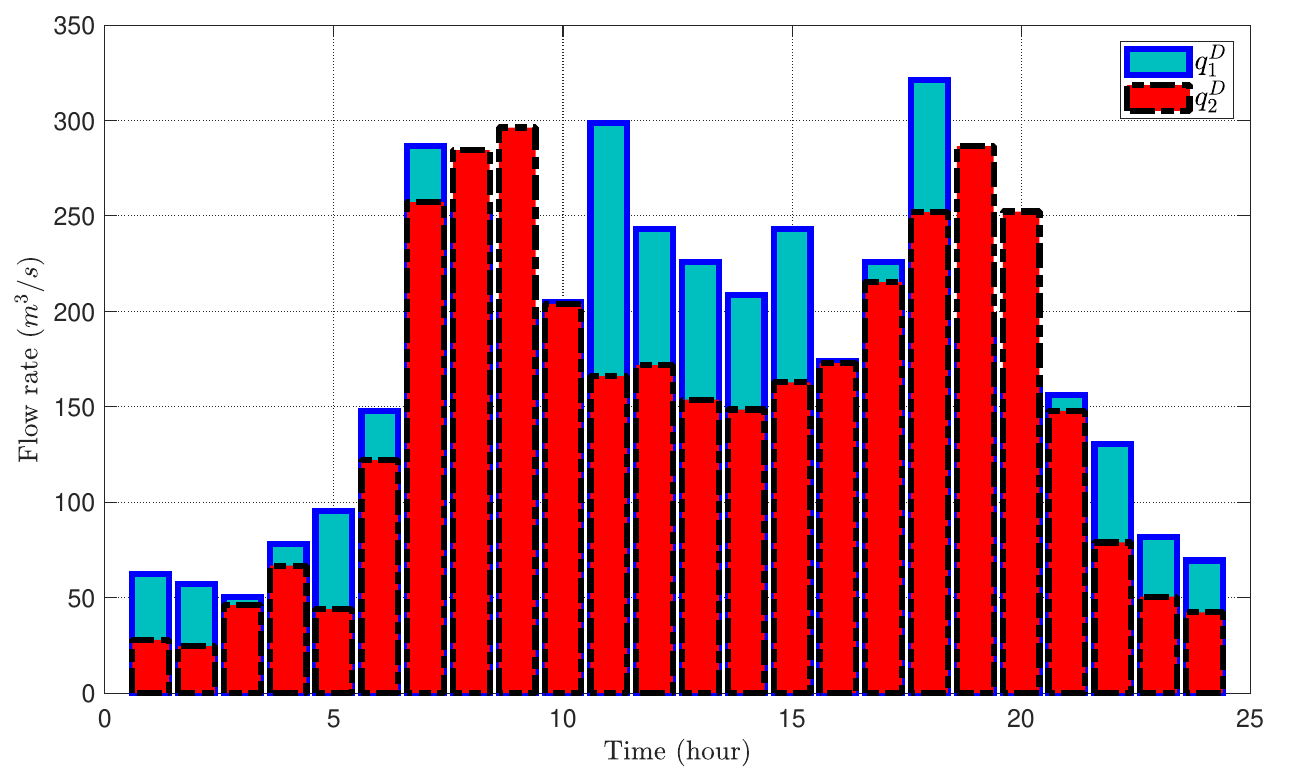}
    \caption{Water demand pattern}
    \label{fig:demand}
\end{figure}

\begin{figure}[t!]
    \centering
    \includegraphics[width=1\columnwidth]{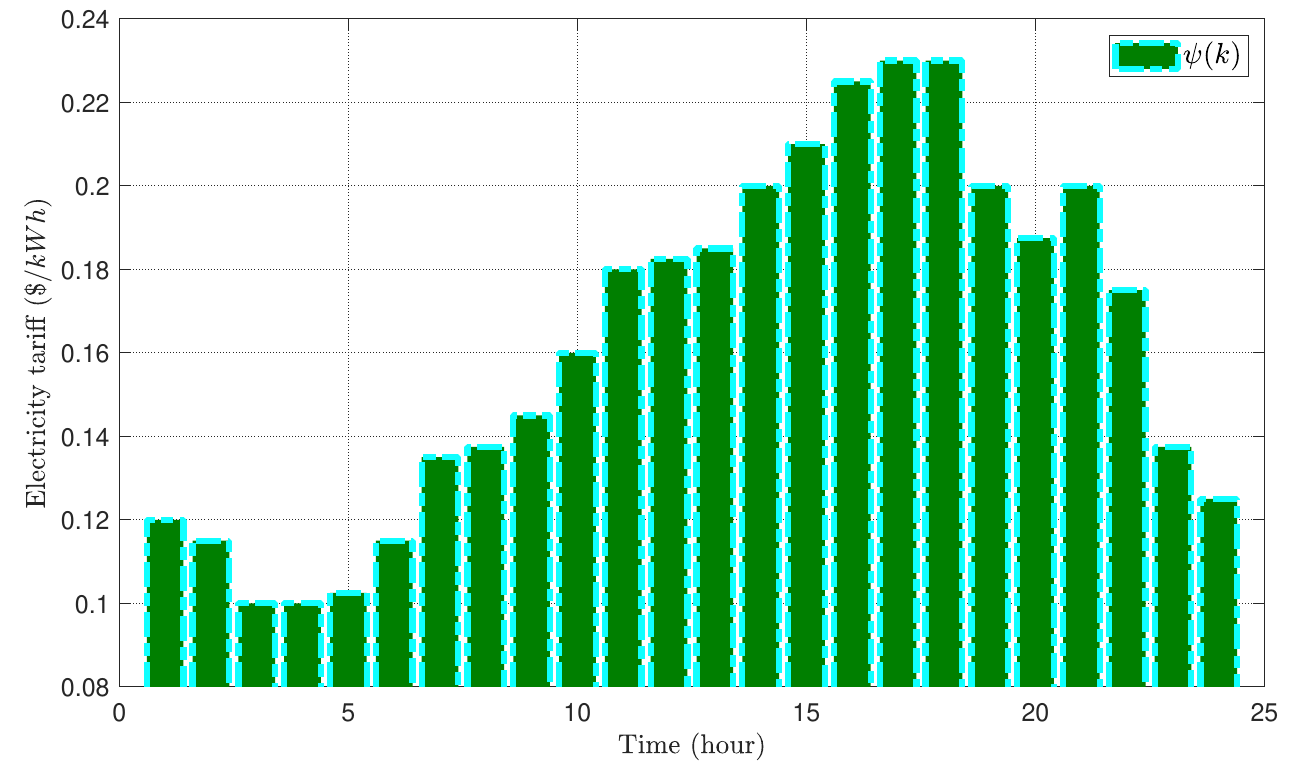}
    \caption{Electricity tariff}
        \label{fig:tariff}
\end{figure}

In this study, water demand is regarded as a known disturbance, depicted in Fig.\ref{fig:demand}. Two water demand patterns are utilized to provide variability in evaluating the effectiveness of the controller. Furthermore, to provide a cost-effective operation, Fig.~\ref{fig:tariff} shows the time-varying electricity tariff which will be embedded in the MPC objective function such that the generated pump scheduling can be operated economically and reliably.


\subsection{Interpolated move-blocking strategy}
In this study, a modified move-blocking strategy is performed by performing regularized linear interpolation to the blocked control inputs aiming at maintaining feasibility while still achieving the computational benefits. Furthermore, the move-blocking strategy is directly applied to the rate of change of the control inputs ($\bm{\Delta u} (k) = \bm{u}(k) - \bm{u}(k-1)$), herein referred to delta input blocking (DIB), where the control rate of change is fixed, instead of fixing the control input variation ($\bm{\hat{u}}$) \cite{cagienard2007move,son2020move}. The general idea of the move-blocking strategy is that in contrast to the optimal delta input variation sequence $\bm{\bm{\hat{u}}}^{\ast} := [\bm{\Delta u}^{\ast}_0 \ \bm{\Delta u}^{\ast}_{1} \ \dots \ \bm{\Delta u}^{\ast}_{N_p-1}] \in \mathbb{R}^{n_uN_p}$, the solution from MPC can be expressed as the optimal reduced control input variation sequence $\bm{\bar{\bm{\hat{\Delta u}}}}^{\ast} := [\bm{\bar{u}}^{\ast}_0 \ \dots \ \bm{\bar{\Delta u}}^{\ast}_{N_c-1}] \in \mathbb{R}^{n_uN_c}$. Notably, to utilize the move-blocking strategy, the selection of the blocking intervals, position, and matrix is also critical, which are described in Definition~\ref{def1:bint}-\ref{def2:bmat}.

\begin{definition}[Blocking interval and position] \label{def1:bint} \mbox{} \par
    Blocking interval refers to a length of time during which a particular control action is held constant. Let $N_c \in \mathbb{N}_{\leq N_p}$ represent the number of the block interval, equivalent to the control horizon and denoted as $N_c\in \mathbb{N}_{\leq N_p}$ (the set of natural numbers less than or equal to $N_p$), where the controller computes $N_c$ moves. The difference with the classical selection of control horizon is that it is specified into a vector of block sizes, $\bm{l} = [l_1 \ l_2, \hdots, l_{N_c}] \in \mathbb{N}$, herein $\bm{l}$ is called the block length. The summation of the vector must match the prediction horizon ($\sum_{i=1}^{N_c} \bm{l} = N_p$). The block length can be increasing or equal distribution. The blocking position set $s:= \{s_1, \hdots, s_{N_c}\}$ in ascending order is closely related to $\bm{l}$ with each element $s_i \in \mathbb{N}_{\leq N_p}$ representing the blocking position and where the block begins. Let $\mathcal{S}$ be the admissible set of $s$.
\end{definition}
\begin{definition}[Blocking matrix]\label{def2:bmat}\mbox{} \\
    Blocking matrix is a transformation matrix to translate the full set of control input sequence into a smaller set according to the predefined blocking intervals. Let $\bm{\bar{M}}\in {\mathbb{R}}^{N_p \times N_c}$ be the blocking matrix that formulates a lower triangular matrix, whose elements are subjected to binary values $\{0,1\}$ \cite{son2021move}. The position of the nonzero component is defined by the admissible set of the blocking position ($s$) and the block length ($\bm{l}$):
    
    \begin{align}
         \bm{\bar{M}} &:= \left[\bm{\bar{t}}_{1}, \hdots,\bm{\bar{t}}_{N_c} \right] \in \mathbb{R}^{N_p \times N_c} \label{blckmatrix}\\
       \bm{\bar{M}}_i &:= \left[\mathbf{0}_{s_i-1};\mathbf{1}_{l_i }; \mathbf{0}_{N_p-s_i} \right] \ \text{for} \ i = 1, \hdots, N_c  \nonumber
    \end{align} 
    where $\bm{\bar{M}}_i$ is the $i$th column of $\bm{\bar{M}}$, $s_i$ is the $i$th component of $s$, $l_i$ is the $i$th element of $\bm{l}$, $\mathbf{0}$ and $\mathbf{1}$ represent a vector or zeros and ones, respectively with length as specified in Eq.~\eqref{blckmatrix}.  
\end{definition}
By definition~\ref{def1:bint}-\ref{def2:bmat}, the blocking matrix in the control input rate of change can be expressed as follows:
\begin{align}
    \bm{\Delta U} &= \bm{\tilde{M_d}} \bm{\Delta \bar{U}} \nonumber\\
    [\bm{\Delta u}_0, \dots, \bm{\Delta u}_{N_p-1}]^T &= (\bm{\bar{M_d}}\otimes \bm{I}_{nu})[\bm{\bar{\Delta u}}_0, \dots, \bm{\bar{\Delta u}}_{\bar{N}-1}]
\label{DIB}
\end{align}  
Furthermore, to perform the least-restrictive move-blocking strategy, let: 
\begin{align}
    \bm{\Delta u}_{b(i)} &= \lambda \bm{\Delta u}_{s(i)} + (1-\lambda)\bm{\Delta u}_{s(i+1)} \label{interpolate}\\ &\quad \forall i \in \{1,2,\hdots,N_c\} \ \forall s \in \mathcal{S}, \ \forall b \notin \mathcal{S}  \nonumber
\end{align}
where $\lambda \in \{0,1\}$ is the interpolation parameter that acts as the weighted average to the interpolation and can be specifically computed from $\lambda:= 1-\frac{b(i)-s(i)}{s(i+1)-s(i)} \ \forall i \in \{1,2,\hdots,N_c\}$. 

\subsection{Final optimal control problem formulation of the flow-based WDS} 
Accordingly, the final optimal control problem formulation is described as:
\begin{subequations} \label{eq:wdsocp}
\begin{align}
    \min_{\mathbf{U}_k, \mathbf{X}_k,\mathbf{\Xi}_k} & J(\bm{u}(k+j),\bm{x}(k+j),\bm{\xi}(k+j)) \label{wds:objfun}\\
    \text{s.t.} \quad & \bm{x}(k+j) = \mathbf{f}_d(\bm{x}(k+j), \bm{u}(k+j),\bm{d}(k+j)), \label{wds:dyn}\\
    & \quad \quad \quad \quad \quad j = 1, 2, \dots, N_p \nonumber \\ 
    &\mathbf{0} = \mathbf{E}\bm{u}(k+j) +\mathbf{\Lambda}\bm{d}(k+j) \quad j = 0, 1, \dots, N_p-1 \label{wds:mass}\\
    &\bm{\Delta u}(k+j) =  \bm{u}(k+j) - \bm{u}(k+j-1) \nonumber \\
    &\quad j = 0, 1, \dots, N_p-1 \\
    &\bm{\Delta U} = \lambda \bm{\Delta u}_{s(i)} + (1-\lambda)\bm{\Delta u}_{s(i+1)}  + \bm{\tilde{T_d}}\bm{\Delta \bar{U}} \label{wds:IDIB1} \\
    & s = \{s_1, \hdots, s_{N_c}\} \in \mathbb{N}_{\leq N_p} \label{wds:IDIB2}\\
    &\lambda \in \{0,1\} \label{wds:IDIB3}\\
    &i = \{1,2,\hdots,N_c\} \label{wds:IDIB4}\\
    & \bm{u}(k+j) \in \mathcal{U}, \quad j = 0, 1, \dots, N_p-1 \label{wds:ubound}\\ 
    & \bm{x}(k+j) \in \mathcal{X}, \quad j = 1, 2, \dots, N_p \label{wds:statebound}\\
    & \bm{\xi}(k+j) \geq 0, \quad j = 0, 1, \dots, N_p-1 \label{wds:slackbound} \\
    & \bm{x}_0 = \mathbf{f}_d(\bm{x}(k), \bm{u}(k),\bm{d}(k))\label{wds:plant}
\end{align}
\end{subequations}
where $\bm{\xi}\in \mathbb{R}^4$ as the slack vector to accommodate the feasibility of the optimal control problem, $\mathbf{U}_k$, $\mathbf{X}_k$, and $\mathbf{\Xi}_k$ are the optimal input, states, and slacks sequence, respectively over the prediction horizon $N_p$ at time-instant $k$. Constraint~\eqref{wds:dyn} represents the dynamics of the WDS detailed in Eq.~\eqref{tank1}. Constraint~\eqref{wds:mass} ensures the node balance following~Eq.~\eqref{tank2}. Constraints \eqref{wds:IDIB1}-\eqref{wds:IDIB4} enforce the interpolated delta input blocking to accelerate the computation time to solve $J(k)$ in~Eq.~\eqref{tank7}.

\section{RESULTS \& DISCUSSION}\label{sec:casestudies}
This section demonstrates time domain validation of utilizing the fast control algorithm, herein referred to as IDIB-MPC, centered on economic-based MPC. To validate the effectiveness of the proposed method the aggregated WDS is operated over a continuous span of 72 hours (3 days). The proposed control algorithm is solved in MATLAB using ``fmincon" nonlinear solver with a solution method specified to sequential quadratic programming (SQP). The models are carried out on an Intel Core CPU i7-6700 processor at 3.40 GHz and 32GB RAM. Mean absolute percentage error (MAPE) is utilized to measure the discrepancy of the states when applying the blocked and non-blocked MPC, expressed as 
\begin{align}
MAPE = \frac{1}{n} \sum_{i=1}^{n} \left| \frac{{x_i - \hat{x}_i}}{{x_i}} \right| \times 100\%  \label{mapeeq}
\end{align}
where, $n$ represents the total number of samples, $x_i$ represents the desired set-point, and $\hat{x}_i$ represents the predicted value. 
\subsection{Economic WDS operation with IDIB-MPC}
\begin{figure*}[t!]
    \centering
    \includegraphics[width = 1\textwidth]{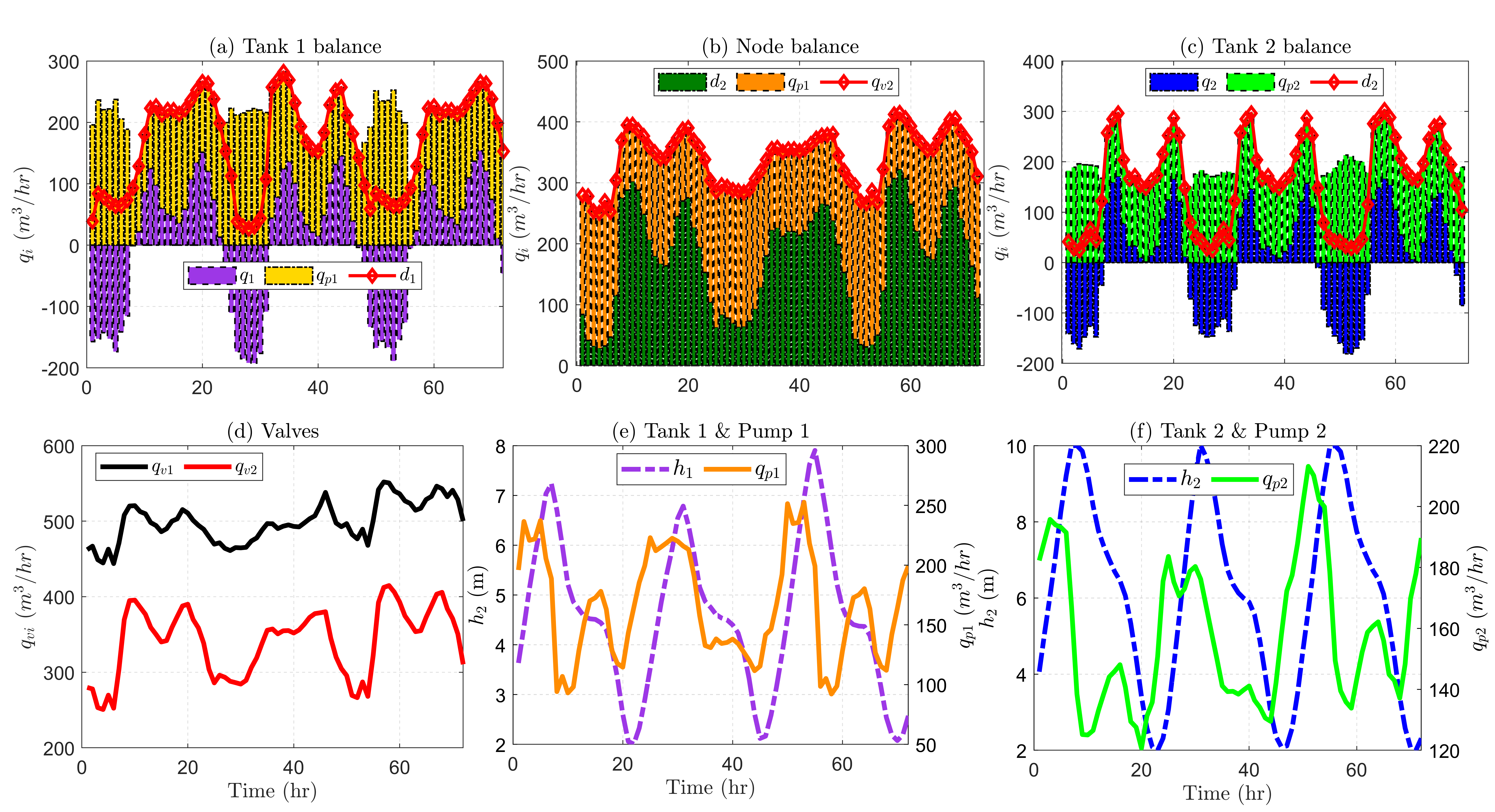}
    \caption{Demand-driven WDS operation under Fast and Economic MPC}
    \label{fig:wtrdemand}
\end{figure*}

In this study, MPC is tailored for cost-effectiveness based on time-varying electricity costs, with an emphasis on the smoothness of the control actions to ensure operational stability. This approach also considers the physical limitations of the hydraulic elements, including the capacities of tanks, pumps, and valves. Fig.~\ref{fig:wtrdemand} shows the dynamic balance of the hydraulic elements over a 72-hour period. Employing a sampling time ($t_s$) of one hour, the prediction horizon ($N_p$) is set at 24 hours. This configuration signifies that with each incremental time step, the controller solves the optimal control sequence and predicts the system's behavior for the next 24 hours while considering the time-varying disturbance (water demand) and the output of the system. Thus endowing the controller with a robust and dependable control mechanism. 

Fig.~\ref{fig:wtrdemand}(a) and (c) display demand management of the proposed WDS using IDIB-MPC. As depicted, water demand is consistently guaranteed throughout the simulation, indicated by an identical graph between the demand and the total flows from the stored water in the tanks and pumps. Notably, there are instances where the flow rates of the tanks ($q_1$ and $q_2$) extend below zero. This visualization aims to better illustrate the ``peak-shaving" behavior of the water demand as a result of the optimal pump scheduling. At peak demands and when electricity tariffs are higher (see Fig.~\ref{fig:tariff}), the stored water is released continuously to meet water demand and reduce pump utilization until the water levels reach the minimum safety thresholds, thereby satisfying the physical limitation and ensuring cost-effective WDS operation.  Furthermore, Fig.~\ref{fig:wtrdemand} (b) exhibits the dynamical flows between the actuators and aggregated demand in a node. It is observed that IDIB-MPC as the controllers adequately adjust the flows to ensure continuous supply despite the demand variation. 

Fig.~\ref{fig:wtrdemand}(d)-(f) further shows the trajectories of the valves, pumps, and water levels of the tanks under cost-effective WDS operation. Similarly, MPC adeptly adjusts the actuators' flow rates to handle the system variations, such as time-varying demand, pump efficiency curve, and electricity prices), while still satisfying the physical limit, thereby optimizing the WDS. A gradual decrease and increase in the flow rates indicate the smoothness of the actuators' operation, achieving the control objectives. 
\subsection{Evaluation on move-blocking strategy}
\begin{figure*}[t!]
    \centering
    \includegraphics[width = 1\textwidth]{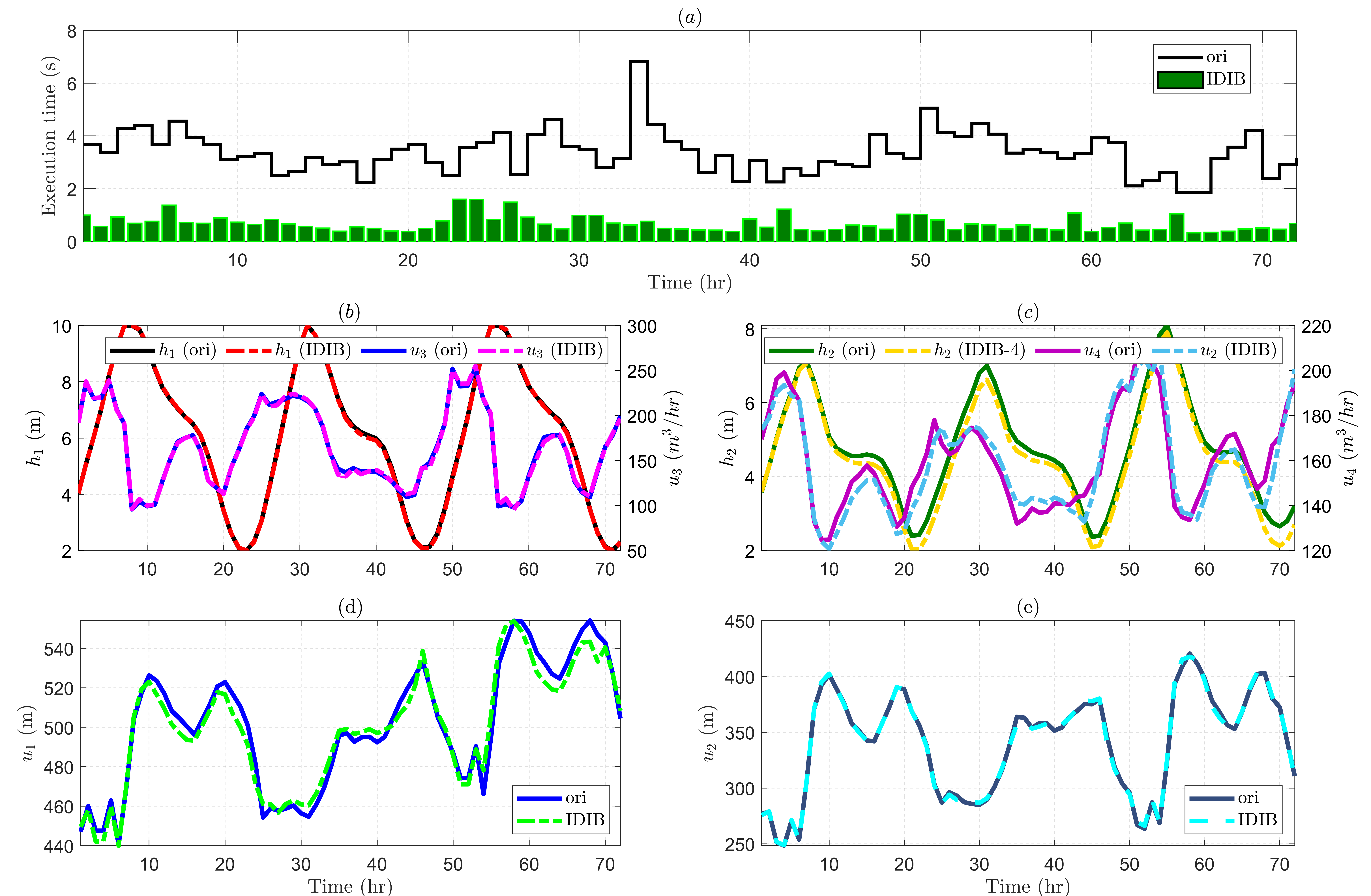}
    \caption{WDS dynamic behavior under fast control algorithm}
    \label{fig:fastmpc}
\end{figure*}

A comparative analysis of an MPC strategy's performance with (IDIB) and without (ori) the application of the least-restrictive move-blocking is exhibited in Fig.~\ref{fig:fastmpc}. In this study, the prediction horizon of $N_p = 24$ and control horizon of $N_c = 6$ area applied with blocking position set $s \in \{1,2,4,5,11,16\}$ and blocking interval $l = [1,2,3,4,5,9]$. The performance metrics and the dynamic behavior of the systems are plotted over 72 hours. 

Fig.~\ref{fig:fastmpc}(a) displays the execution time trajectory for each MPC optimization cycle. It is clearly illustrated that the proposed method consistently requires faster time to solve the optimization problem in MPC throughout the simulation. The proposed method is able to significantly reduce the executing time with an average of 80\%, thereby, suggesting a computationally efficient control. 

The move-blocking strategy, while effective, has inherent limitations with its tendency to deviate from optimal control inputs. As a result, a piecewise linear function is utilized for the blocked inputs to ensure the control inputs remain within boundaries. The efficacy of this approach is illustrated in Fig.~\ref{fig:fastmpc}(b)-(e). The figures present a comparative analysis of system variables under both MPC strategies (with and without move-blocking). Both MPC strategies exhibit similar trajectories, as evidenced by the overlapping lines. This similarity in trajectories indicates that the proposed method while reducing the computation time, does not compromise the optimality of the optimization process.   

\subsection{Quantifying Performance of the Interpolated Delta-Input Blocking}
\begin{figure}[t!]
    \centering
    \includegraphics[width = 1\columnwidth]{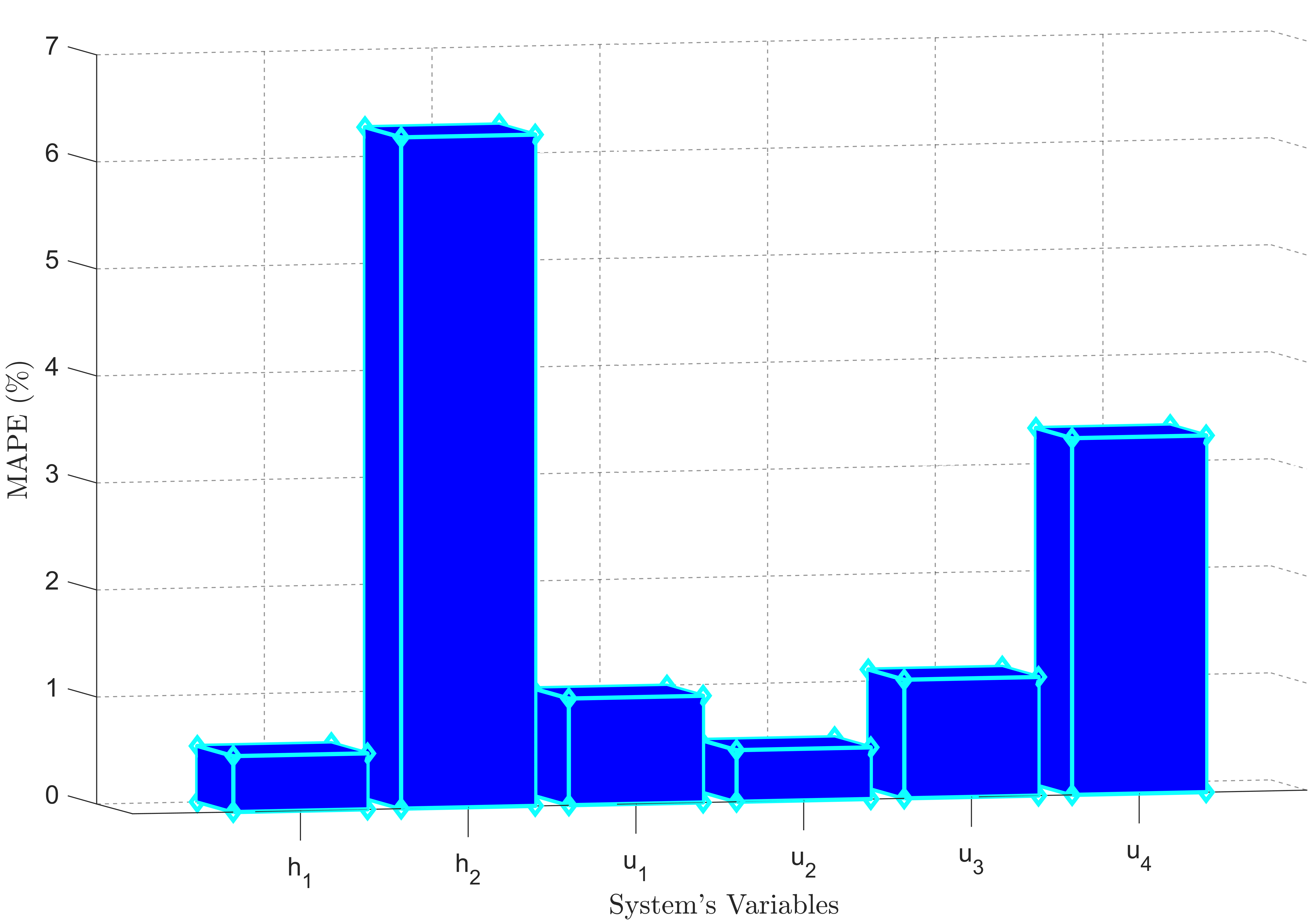}
    \caption{Mean absolute percentage error ($\%$).}
    \label{fig:mapeeror}
\end{figure}

Fig.~\ref{fig:mapeeror} illustrates the impact of the fast algorithm on the WDS via MAPE computation for all system variables, using Eq.~\eqref{mapeeq}. A low MAPE consistently below 10\% is observed for all system variables, indicating the high accuracy of the interpolated-delta input blocking. Flow rates of valve 2 ($q_{v2}$) were observed to have the lowest MAPE of 0.5\%, while water levels of tank 2 were observed to have the highest MAPE of 6.3\%. This indicates that in addition to reducing MPC computation with an average of 80\%, the proposed method can still maintain the performance of non-blocked MPC. 
\section{Conclusion}\label{sec:conclusion}
Model predictive control (MPC) has been the preferred control in water distribution systems (WDSs) as it includes explicit constraints, provides feedback control to correct the control actions, and provides future prediction of the system's output. However, MPC is limited by its high computation costs to yield optimal control solutions, specifically for multiple input and multiple output (MIMO) systems and nonlinear control in large-scale WDSs. 

This study aims to overcome this limitation by applying a fast computation algorithm that utilizes an interpolated move-blocking strategy for controlling WDSs via MPC. The proposed method is validated through the control of an aggregated WDS, focusing on demand-driven and cost-effective operation. Multi-objective functions and time-varying penalty weights that depend on electricity cost and actuators' performance curve are considered. 

A least-restrictive algorithm of the move blocking is developed via piecewise linear delta-input blocking, and predefined blocking interval and position distribution are implemented. The results show that the proposed method has significantly reduced the online computation by an average of 80\% with consistent optimality convergence. An identical trajectory between the full degrees of freedom (DoF) and the blocked inputs as well as the states of the systems are observed with MAPE consistently below 10\%, respectively. In light of these findings, it is confirmed that MPC with the least-restrictive move-blocking strategy provides an alternate way to ameliorate the computational burden in MPC, especially for controlling large-scale water distribution systems in real-time.

\bibliographystyle{IEEEtran}
\bibliography{IEEEabrv,ewribib2}

\end{document}